\documentclass[12pt]{article}
\usepackage{amsmath}
\usepackage{bm}
\usepackage{color}
\usepackage{braket}
\usepackage{caption}
\usepackage{subcaption}
\usepackage{graphicx}

\begin{document}
\begin{center}
\begin{large}
{\bf Entanglement of multi-qubit states representing directed networks and its detection with quantum computing}
\end{large}
\end{center}

\centerline {Kh. P. Gnatenko \footnote{E-Mail address: khrystyna.gnatenko@gmail.com}}
\medskip
\centerline {\small \it Ivan Franko National University of Lviv,}
\centerline {\small \it Professor Ivan Vakarchuk Department for Theoretical Physics,}
\centerline {\small \it 12 Drahomanov St., Lviv, 79005, Ukraine}

\abstract{We consider quantum graph states that can be mapped to directed weighted graphs, also known as directed networks. The geometric measure of entanglement of the states is calculated for the quantum graph states corresponding to arbitrary graphs. We find relationships between the entanglement and the properties of the corresponding graphs. Namely, we obtain that the geometric measure of entanglement of a qubit with other qubits in the graph state is related to the weights of ingoing and outgoing arcs with respect to the vertex representing the qubit, outdegree and indegree of the corresponding vertex in the graph. For unweighted and undirected graphs, the entanglement depends on the degree of the corresponding vertex.  Quantum protocol for quantifying of the entanglement of the quantum graph states is constructed. As an example, a quantum graph state corresponding to a chain is examined, and the entanglement of the state is calculated on AerSimulator.

Keywords: quantum graph states; geometric measure of entanglement; directed weighted graph.
}

\section{Introduction}

Quantum graph states have been recently studied intensively \cite{ Markham,Wang, Mooney, Schlingemann, Bell, Mazurek, Shettell, Hein, Guhne, Qian, Vesperini1, Vesperini2}, and widely used in quantum information and quantum programming, for instance, in quantum error correction algorithms \cite{Schlingemann, Bell, Mazurek}, quantum cryptography \cite{Markham, Qian}, quantum machine learning \cite{Gao, Zoufal}, and many other applications. These states can be represented by graphs. Therefore, based on studies of the quantum properties of quantum graph states, the properties of graphs can be detected and vice versa (see, for instance, \cite{Laba22}). Additionally, quantum graph states possess an important property, such as entanglement.

Entanglement of quantum states is one of the most important resources in quantum computing, reflected in particle correlations and the non-factorization of a quantum state \cite{Horodecki, Feynman, Ekert, Bennett, Lloyd, Bouwmeester, Raussendorf, Buluta, Horodecki1, Shi, Llewellyn, Huang, Yin, Jennewein, Karlsson}. Many papers are devoted to the study of the entanglement of quantum states, both analytically and through quantum computing (see, for instance, \cite{Horodecki, Shimony, Behera, Scott, Horodecki1, Torrico, Sheng, Samar, Kuzmak,  Gnatenko, Kuzmak2, Vesperini1, Vesperini2, Vesperini3} and references therein).

In recent papers, quantum graph states constructed with the action of controlled-Z gates on the state $\ket{+++...+}$ were examined (see, for example, \cite{Wang, Mooney, Alba, Mezher, Akhound, Haddadi, Cabello}). These states are multi-qubit states that can be related to undirected graphs $G(V,E)$. The qubits are represented by the vertices of a graph, edges are represented with the actions of controlled -Z gates. Namely, we have $\ket{\psi_G} = \prod_{(i,j) \in E} CZ_{ij}\ket{+}^{\otimes V}$, where $CZ_{ij}$ is the controlled-Z gate that acts on qubits $q[i]$ and $q[j]$. A more general case of graph states, namely quantum graph states constructed with the action of controlled phase gates $CP_{ij}(\phi)$ on a separable multi-qubit state, was examined in \cite{Susulovska, Susulovska1}, and the entanglement of the states was calculated both analytically and with quantum programming. Additionally, quantum graph states of spin systems with Ising interaction were considered \cite{Gnatenko}. The entanglement of these states was studied \cite{Gnatenko, Laba22}. In \cite{Gnatenko}, the geometric properties of the states were examined, and the relation of these properties with the number of triangles, squares, and edges was found \cite{Laba22}.

To distinguish entangled states, different measures of entanglement have been introduced. In the present paper, we consider the geometric measure of entanglement, which was proposed in \cite{Shimony} as the minimal squared Fubini-Study distance between an entangled state $\vert\psi\rangle$ and a set of separable nonentangled states. This definition, proposed by Shimony, reads:
\begin{eqnarray}
E(\vert\psi\rangle)=\min_{\vert\psi_s\rangle}d_{FS}^2 (\vert\psi\rangle, \vert\psi_s\rangle),\\
d_{FS}(\vert\psi\rangle, \vert\psi_s\rangle)=\sqrt{1-|\langle\psi|\psi_s\rangle|^2},
\end{eqnarray}
where $\vert\psi\rangle$ is an entangled state and $\vert\psi_s\rangle$ is a set of separable nonentangled states.

In this paper, we consider quantum graph states constructed by the action of two-qubit gate  $RXX(\phi_{ij})=\exp(-i\phi_{ij}\sigma^x_i\sigma^x_j/2)$ gates ($\sigma^x_i$, $\sigma^x_j$ are Pauli matrices corresponding to qubits $q[i]$ and $q[j]$) on an arbitrary separable multi-qubit quantum state. These states are related to directed and weighted graphs (directed networks). In the general case of a quantum state corresponding to an arbitrary directed network, we have found the analytical result for the geometric measure of entanglement of the state. It is shown that the entanglement is related to the graph properties; namely, the entanglement of a qubit with other qubits in the state is related to the weights of the arcs, outdegree, and indegree of the corresponding vertex in the graph.

The paper is organized as follows. In Section 2, the geometric measure of entanglement of quantum graph states that correspond to directed weighted graphs is studied. The relation of the entanglement to the graph properties is presented. In Section 3  quantum protocol for the detection of the entanglement is constructed. The results of the calculation of the entanglement of the  quantum graph states on AerSimulator \cite{kk} are presented. Conclusions are drawn in Section 4.

\section{Geometric measure of entanglement of quantum states corresponding to directed networks}

We consider multi-qubit quantum states in the following form
\begin{eqnarray}
\ket{\psi_G}=\prod_{(i,j) \in A} RXX_{ij}(\phi_{ij}) \ket{\psi_0}. \label{eq:2:1}
\end{eqnarray}
The states are prepared with the action of two-qubit gates $RXX_{ij}(\phi_{ij})$
on the initial state $\ket{\psi_0}$, which is an arbitrary separable state of $V$ qubits
\begin{eqnarray}
\ket{\psi_0}=\prod_{k \in V} \ket{\psi(\alpha_k,\theta_k)}_k,\\
\ket{\psi(\alpha_k,\theta_k)}_k = \cos \frac{\theta_k}{2} \ket{0}_k + e^{i\alpha_k} \sin \frac{\theta_k}{2} \ket{1}_k = e^{i\frac{\alpha_k}{2}}RZ(\alpha_k)RY(\theta_k)\ket 0_k. \label{eq:2:2}
\end{eqnarray}
Here $RZ(\alpha_k)=\exp\left(-i{\alpha{k}}\sigma^z_k/2\right)$, $RY(\theta_k)=\exp\left(-i{\theta{k}}\sigma^y_k/2\right)$ are rotational gates, $\sigma^{y}_k$, $\sigma^{z}_k$ are Pauli matrices corresponding to qubit $q[k]$. The state $\ket{\psi_G}$ can be represented by the directed and weighted graph $G(V,A)$ ($V$ are graph vertices and $A$ are arcs) with the adjacency matrix with parameters $\phi_{ij}$. The $G(V,A)$ does not contain self-loops (arcs that connect vertices to themselves), namely $\phi_{ii}=0$.

According to the results of the paper \cite{Samar}, the geometric measure of entanglement of a spin with other spins in a system is related to its mean value. Namely, it was found that the geometric measure of entanglement of a spin $1/2$ (a qubit) with another spin system (system of qubits) in a state $\vert\psi\rangle$ reads
\begin{eqnarray}
E(\vert\psi\rangle)=\frac{1}{2}\left(1-|\langle{\bm \sigma}\rangle|\right)=
\frac{1}{2}\left(1-\sqrt{\langle\sigma^x\rangle^2+\langle\sigma^y\rangle^2+\langle\sigma^z\rangle^2}\right),\label{ent}
\end{eqnarray}
where $\langle{\bm \sigma}\rangle$ is the mean value of the spin in the state $\ket\psi$.

Let us calculate the entanglement of qubit $q[k]$ with other qubits in the quantum state $(\ref{eq:2:1})$. For the mean value $\braket{\sigma^x_k}$ we have
\begin{eqnarray}
\braket{\sigma^x_k} = \bra{\psi_G} \sigma^x_k \ket{\psi_G} =
{}_k\bra{\psi(\alpha_k,\theta_k)} \sigma^x_k \ket{\psi(\alpha_k,\theta_k)}_k = \cos \alpha_k \sin \theta_k \label{sx}
\end{eqnarray}

The analytical result for the mean value $\braket{\sigma^y_k}$ reads
\begin{eqnarray}
\braket{\sigma^y_k} = \bra{\psi_G} \sigma^y_k \ket{\psi_G} = \nonumber\\=\bra{\psi_0} \prod_{m \in U} \prod_{m \in W} RXX^+_{mk}(2\phi_{mk}) RXX^+_{kn}(2\phi_{kn}) \sigma^y_k \ket{\psi_0} = 2\Re z \label{sy}
\end{eqnarray}
Here $U$ is the set of vertices that are connected with vertex $k$ by the arcs incident to $k$ (arc $mk$ with $m \in U$ leaves $m$ and enters $k$). Also, we introduced the notation $W$ for the set of vertices that are connected with vertex $k$ by the arcs that leave $k$. Arc $kn$ with $n \in W$ leaves $k$ and enters $n$. Number $z$ is the complex number as follows
\begin{eqnarray}
z=\frac{1}{2}\left(\sin \alpha_k \sin \theta_k + i \cos \theta_k\right) \times \nonumber\\
\mathop{\prod_{m \in U} \prod_{n \in W}}\limits_{m\neq n} \left(\cos \phi_{mk} + i \sin \phi_{mk} \cos \alpha_m \sin \theta_m\right) \left(\cos \phi_{kn} + i \sin \phi_{kn} \cos \alpha_n \sin \theta_n\right) \times \nonumber\\ \prod_{l \in B} \left(\cos (\phi_{lk} + \phi_{kl}) + i \sin (\phi_{lk} + \phi_{kl}) \cos \alpha_l \sin \theta_l\right).
\end{eqnarray}
Here $B$ is the set of vertices that are connected with vertex $k$ with two arcs with opposite directions (bidirected arcs $lk$, $kl$, $l \in B$ connect $l$ and $k$).
To obtain (\ref{sy}), we take into account that
\begin{eqnarray}
RXX^+_{mk}(\phi_{mk}) RXX^+_{kn}(\phi_{kn}) \sigma^{y}_k RXX_{mk}(\phi_{mk}) RXX_{kn}(\phi_{kn})=\nonumber\\ = RXX^+_{mk}(2\phi_{mk}) RXX^+_{kn}(2\phi_{kn}) \sigma^{y}_k,
\end{eqnarray}
this is  due to the anticommutation of operators $\sigma^x_m\sigma^x_k$, $\sigma^x_k\sigma^x_n$,   $\sigma^{y}_k$.

Similarly, for $\sigma^z_k$ we find
\begin{eqnarray}
\braket{\sigma^z_k} = \bra{\psi_G} \sigma^z_k \ket{\psi_G} = \nonumber\\ =
\bra{\psi_0} \prod_{m \in U} \prod_{m \in W} RXX^+_{mk}(2\phi_{mk}) RXX^+_{kn}(2\phi_{kn}) \sigma^z_k \ket{\psi_0} = 2\Im z \label{sy2}
\end{eqnarray}

On the basis of the results for the mean values, the geometric measure of entanglement of qubit $k$ in the graph state $\ket{\psi_G}$ reads
\begin{eqnarray}
E_k(\ket{\psi_G}) = \frac{1}{2} - \frac{1}{2} \left[ \cos^2 \alpha_k \sin^2 \theta_k +  \left( \cos^2 \theta_k + \sin^2 \alpha_k \sin^2 \theta_k \right) \times \right. \nonumber\\ \left.
\mathop{\prod_{m \in U} \prod_{n \in W}}\limits_{m\neq n} \left( \cos^2 \phi_{mk} + \sin^2 \phi_{mk} \cos^2 \alpha_m \sin^2 \theta_m \right) \times\right. \nonumber\\ \left. \left( \cos^2 \phi_{kn} + \sin^2 \phi_{kn} \cos^2 \alpha_n \sin^2 \theta_n \right) \times \right. \nonumber\\ \left. \prod_{l \in B} \left( \cos^2 (\phi_{lk} + \phi_{kl}) + \sin^2 (\phi_{lk} + \phi_{kl}) \cos^2 \alpha_l \sin^2 \theta_l \right) \right]^{\frac{1}{2}}, \label{E_gen}
\end{eqnarray}

In the particular case of $\phi_{mk} = \phi^{(in)}_k$, $\phi_{kn} = \phi^{(out)}_k$, namely when the weights of the arcs that leave $k$ are the same, and the weights of the arcs that enter $k$ are equal, and in the case of absence of bidirected arcs corresponding to vertex $k$, the expression for the entanglement is reduced to
\begin{eqnarray}
E_k(\ket{\psi_G})=\frac{1}{2}-\frac{1}{2}\left[\cos^2 \alpha \sin^2 \theta +\left(\cos^2 \theta+\sin^2 \alpha\sin^2\theta\right)\times \right.\nonumber\\ \left. \times
\left(\cos^2 \phi^{(in)}_k+ \sin^2\phi^{(in)}_k\cos^2\alpha\sin^2\theta\right)^{n^{(in)}_k} \times \right.\nonumber\\ \left. \left(\cos^2 \phi^{(out)}_k+ \sin^2\phi^{(out)}_k\cos^2\alpha\sin^2\theta\right)^{n^{(out)}_k}\right]^{\frac{1}{2}}.\label{entkk}
\end{eqnarray}
Here we also considered the initial state  $\ket{\psi_0}$  to be separable state in which $V$ qubits are in arbitrary states with the same parameters  $\ket{\psi(\alpha_i,\theta_i)}_i =\ket{\psi(\alpha,\theta)}_i$,   $\theta_i=\theta$, $\alpha_i=\alpha$.
In (\ref{entkk}) parameter $n^{(out)}_k$ is the  outdegree of the vertex $k$ (number of arcs that leave $k$), $n^{(in)}_k$  indegree of the vertex $k$ (number of arcs that enter $k$).

Let us also examine the case of an unweighted and undirected graph $G(V,E)$ that can be represented with the state (\ref{eq:2:1}) with $\phi_{ij}=\phi$, namely

\begin{eqnarray}
\ket{\tilde{\psi}_G}=\prod_{(i,j) \in E} RXX_{ij}(\phi) \prod_{k \in V} \ket{\psi(\alpha,\theta)}_k,
\end{eqnarray}
here $E$ are edges of the graph $G(V,E)$, we also consider $\alpha_i=\alpha$, $\theta_i=\theta$.
For the entanglement of qubit $q[k]$, we have that it is related to the degree of vertex $n_k$ representing it in the graph
\begin{eqnarray}
E_k(\ket{\tilde{\psi}_G})=\frac{1}{2}-\frac{1}{2}\left(\cos^2 \alpha \sin^2 \theta +\left(\cos^2 \theta+\sin^2 \alpha\sin^2\theta\right)\times \right.\nonumber\\ \left. \times
\left(\cos^2 \phi+ \sin^2\phi\cos^2\alpha\sin^2\theta\right)^{n_k}\right)^{\frac{1}{2}}\label{par2}
\end{eqnarray}

In the particular case $\theta=0$, $\alpha=0$, the initial state $\ket{\psi_0}$ reduces to $\ket{0}^{\otimes V}$, and on the basis of (\ref{eq:2:1}), we obtain
\begin{eqnarray}
E_k(\tilde{\psi}_G)=\frac{1}{2}(1-|\cos^{n_k}\phi|),
\end{eqnarray}
which is in agreement with the result of the paper \cite{Gnatenko}, where the entanglement of graph states $\ket{\tilde{\psi}_G}=\prod_{(i,j) \in E} RXX_{ij}(\phi) \ket{0..0}$ was examined.

 It is important to note that the entanglement of the graph state is related to the parameters of the corresponding graph. In the general case, we have that the entanglement of qubit $k$ with other qubits in the graph state is determined by the weights of ingoing and outgoing arcs with respect to vertex $k$, and by the sum of weights of antiparallel arcs $lk$, $kl$.

In the particular case of equal weights of arcs that enter $k$ ($\phi_{mk}=\phi^{(in)}_k$), and the weights of arcs leaving $k$ are the same ($\phi_{kn}=\phi^{(out)}_k$) and the bidirected arcs in the graph are absent, the entanglement of qubit $q[k]$ in the graph state (\ref{eq:2:1}) is related to the outdegree and indegree of the corresponding vertex $k$. For quantum state corresponding to an unweighted and undirected graph, the entanglement of $q[k]$ in the state depends on the degree of vertex $k$ that represents the qubit in the corresponding graph.

So, by studying the entanglement of the quantum graph states on a quantum device, these parameters can be determined.

\section{Quantifying the geometric measure of entanglement of the quantum graph states with quantum computing}

\begin{figure}[h!]
		\centering
	\includegraphics[scale=0.6]{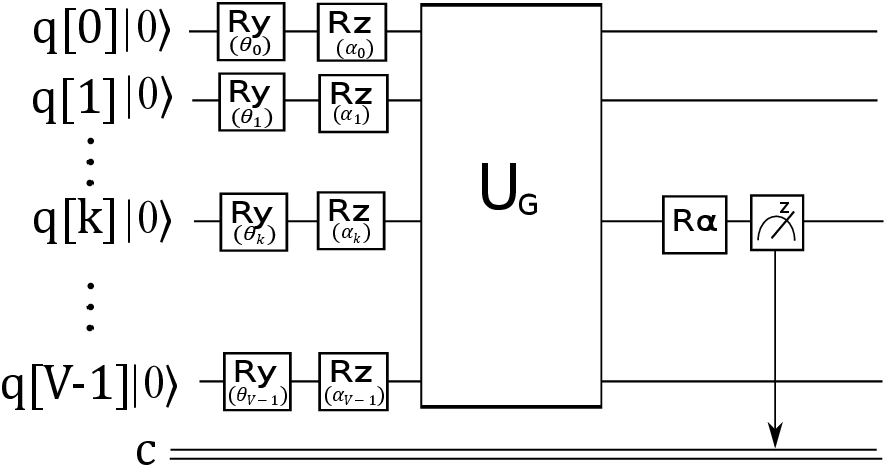}
		\caption{ Quantum protocol for studies of the geometric measure of entanglement of qubit $q[k]$ with other qubits in quantum graph state (\ref{eq:2:1}) on the basis of detection of the mean value of spin $\langle{\bm \sigma}_k\rangle$ .}
		\label{fig:1}
\end{figure}

To quantify the geometric measure of entanglement of a qubit  $q[k]$ (a spin-$1/2$ $k$) with other qubits (other spins) in  the quantum states corresponding to directed and weighted graphs one has to prepare the states and then detect the mean value of spin operator $\langle{\bm \sigma}_k\rangle$ corresponding to the spin.  The mean value of $\sigma^z_k$ operator can be easily detected on the basis of the results of measurement in the standard basis. We have $\langle\sigma^z_k \rangle=\langle \psi_G \vert \sigma^z_k\vert\psi_G\rangle=\vert\langle\tilde\psi_G\vert 0 \rangle\vert^2-\vert\langle\tilde\psi_G\vert 1 \rangle\vert^2$. To find the mean values $\langle\sigma^x \rangle$, $\langle\sigma^y\rangle$, one can use the identities  $\sigma^x_k=\exp(-i\pi\sigma^y/4)\sigma^z_k\exp(i\pi\sigma^y/4)$, $\sigma^y_k=\exp(i\pi\sigma^x/4)\sigma^z_k\exp(-i\pi\sigma^x/4)$.
So, to obtain  $\langle\sigma^x \rangle$  we have to use  $RY_k(-\pi/2)$ before the measurement in the standard basis. To find  $\langle\sigma^y_k \rangle$ one has to apply $RX_k(\pi/2)$ before the measurement. Then the results for the mean values read $\langle\sigma^x_k \rangle=\vert \langle \tilde\psi^{ry}_G \vert 0 \rangle \vert^2-\vert \langle \tilde\psi_G^{ry} \vert 1 \rangle \vert^2$, $\langle\sigma^y_k \rangle=\vert \langle \tilde\psi^{rx}_G \vert 0 \rangle \vert^2-\vert \langle \tilde\psi_G^{rx} \vert 1 \rangle \vert^2$,
 where $\vert\tilde\psi^{ry}_G\rangle=RY_k(-\pi/2)\vert\psi_G\rangle$, $\vert\tilde\psi^{rx}_G\rangle=RX_k(\pi/2)\vert\psi_G\rangle$.
The quantum protocol for the detection of the entanglement of a qubit $q[k]$ with other qubits in the quantum graph state (\ref{eq:2:1}) is presented in the Fig. \ref{fig:1}.

In the quantum protocol shown in Fig. \ref{fig:1}, we use the notation $U_G$ for the operator $U_G=\prod_{(i,j) \in A} RXX_{ij}(\phi_{ij})$. Also, $R_\alpha$ is the rotational gate, which, in the case of detecting the mean value $\langle\sigma^x_k \rangle$, reads $R_\alpha=RY(-\pi/2)$. For studying $\langle\sigma^y_k \rangle$, gate $R_{\alpha}$ has the form $R_\alpha=RX(\pi/2)$. In the case of quantifying $\langle\sigma^z_k \rangle$, this gate is just the identity gate $R_\alpha=I$.

\begin{figure}[h!]
		\centering
	\includegraphics[scale=0.8]{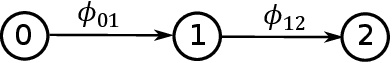}
		\caption{Directed and weighted graph corresponding to a chain. Arcs $01$, $12$ are characterized by weights $\phi_{01}$, $\phi_{12}$.}
		\label{fig:2}
	\end{figure}

Let us consider a particular case of a quantum graph state and study the dependence of the entanglement on the parameters of the quantum state. Let us examine a quantum graph state corresponding to a chain (see Fig. \ref{fig:2}).

\begin{figure}[h!]
		\centering
	\includegraphics[scale=0.6]{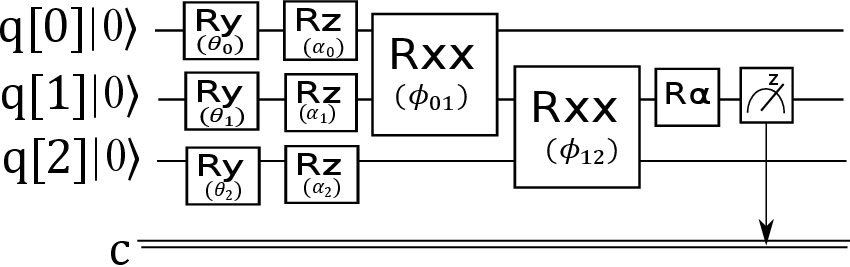}
		\caption{Quantum protocol for studies of the entanglement of qubit $q[1]$ with other qubits in quantum state (\ref{graph_ch}).}
		\label{fig:3}
	\end{figure}

\begin{figure}[h!]
\begin{center}
\subcaptionbox{\label{ff1}}{\includegraphics[scale=0.7, angle=0.0, clip]{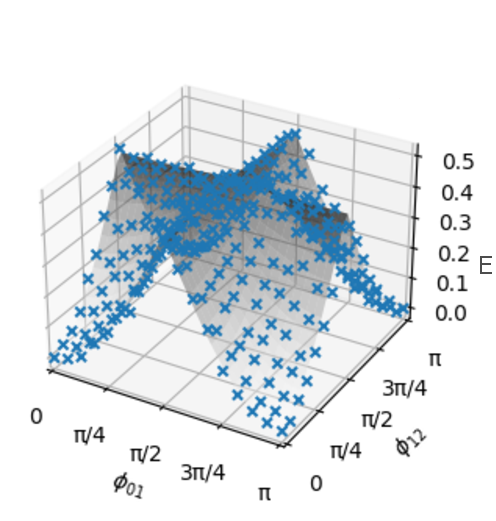}}
\hspace{1cm}
\subcaptionbox{\label{ff3}}{\includegraphics[scale=0.7, angle=0.0, clip]{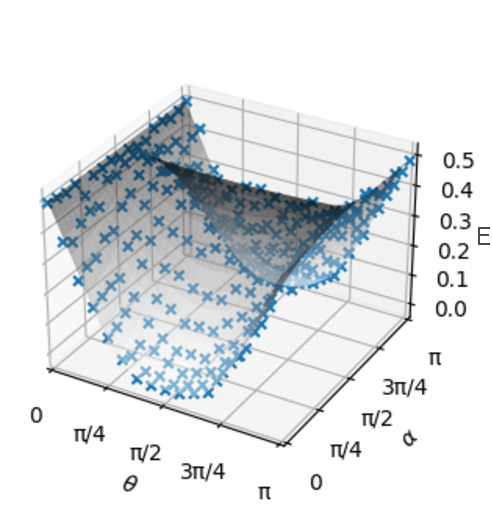}}
\caption{Entanglement of qubit $q[1]$ with other qubits in quantum graph state (\ref{graph_ch}) (a) for $\alpha_i=0$, $\theta_i=0$, $i=(0,1,2)$ and different values of $\phi_{01}$, $\phi_{12}$; (b) for $\phi_{01}=\phi_{12}=\pi/2$, $\alpha_i=\alpha$, $\theta_i=\theta$. Results of calculations on AerSimulator are marked with crosses, the surface corresponds to the analytical results}
		\label{fig:4}
\end{center}
	\end{figure}

This state reads
 \begin{eqnarray}
\ket{\psi_G}=  RXX_{01}(\phi_{01}) RXX_{12}(\phi_{12})\prod^2_{i=0}RZ_i(\alpha_i)RY_i(\theta_i)\ket{000}.\label{graph_ch}
\end{eqnarray}
Quantum protocol for studies of the entanglement of $q[1]$ in this state is presented on Fig. \ref{fig:3}.
We ran this protocol on AerSimulator with 1024 shots. The results of the calculations are presented in Fig. \ref{fig:4}. We considered two cases. The first case is $\alpha_i=0$, $\theta_i=0$, $i=(0,1,2)$, with $\phi_{01}$ and $\phi_{12}$ changing from $0$ to $\pi$ in steps of $\pi/16$ (see Fig. \ref{fig:4}(a)). The second one is characterized by $\phi_{01}=\phi_{12}=\pi/2$, $\alpha_i=\alpha$, and $\theta_i=\theta$ (see Fig. \ref{fig:4}(b)). The analytical results for these cases are
\begin{eqnarray}
E(\phi_{01},\phi_{12})=\frac{1}{2}\left(1-|\cos\phi_{01}\cos\phi_{12}|\right),\\
E(\alpha, \theta)=\frac{1}{2}\left(1-|\cos\alpha\sin\theta|\sqrt{1+\cos^2\alpha\sin^2\theta(\cos^2\theta+\sin^2\alpha\sin^2\theta)}\right),
\end{eqnarray}
respectively.

For the case shown in Fig. \ref{fig:4}(a), the maximum value of entanglement $E=1/2$ is achieved for $\phi_{01}=\pi/2+\pi n$ or $\phi_{12}=\pi/2+\pi n$,  $n \in Z$. For the case presented  in Fig. \ref{fig:4} (b), the maximum entanglement of $q[1]$ with other qubits is achieved for $\theta=\pi n$ and $\alpha=\pi/2+\pi n$,  $n \in Z$

\begin{figure}[h!]
\begin{center}
\subcaptionbox{\label{ff1}}{\includegraphics[scale=0.7, angle=0.0, clip]{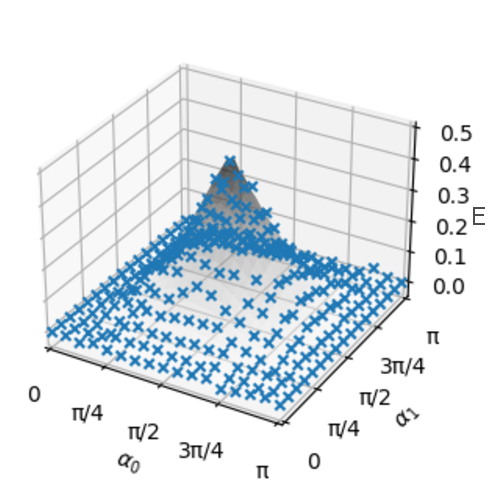}}
\hspace{1cm}
\subcaptionbox{\label{ff3}}{\includegraphics[scale=0.7, angle=0.0, clip]{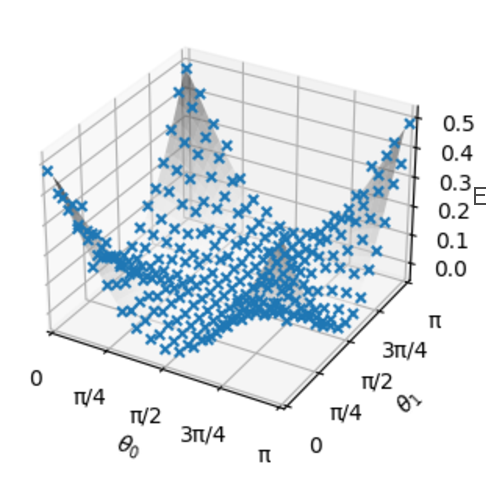}}
\caption{Entanglement of qubit $q[0]$ with other qubits in quantum graph state (\ref{graph_ch}) (a) for $\phi_{01}=\phi_{12}=\pi/2$ and different values of $\alpha_0$, $\alpha_1$; (b) for $\alpha_{i}=0$, $\phi_{01}=\phi_{12}=\pi/2$  and different values of $\theta_0$, $\theta_1$. Results of calculations on  AerSimulator are marked with crosses, the surface corresponds to the analytical result}
		\label{fig:5}
\end{center}
	\end{figure}

Additionally, the entanglement of qubit $q[0]$ with other qubits in the state (\ref{graph_ch}) was examined analytically and using the AerSimulator (with 1024 shots). The results of the calculations on AerSimulator for $\phi_{01}=\phi_{12}=\pi/2$ and $\alpha_0$, $\alpha_1$ changing from $0$ to $\pi$ in steps of $\pi/16$, as well as the analytical result $E(\alpha_0, \alpha_1)=(1-\sqrt{\cos^2 \alpha_0+\sin^2\alpha_0 \cos^2\alpha_1})/2$, are presented in (\ref{fig:5}) (a).

The geometric measure of entanglement of $q[0]$ in the case of $\alpha_i=0$, $\phi_{01}=\phi_{12}=\pi/2$, and different values of $\theta_0$ and $\theta_1$ changing from $0$ to $\pi$ in steps of $\pi/16$ is shown in (\ref{fig:5}) (b). Analytical result in this case reads $E(\theta_0, \theta_1)=(1-\sqrt{\sin^2 \theta_0+\cos^2\theta_0 \sin^2\theta_1})/2$.

 The maximal value of entanglement of qubit $q[0]$ with other qubits is achived for   $\alpha_0=\pi/2+\pi n$ and $\alpha_1=\pi/2+\pi n$ (case (a)) and  $\theta_0=\pi n$, $\theta_1=\pi n$, $n\in Z$ (case (b)).

 \section{Conclusions}

Multi-qubit quantum graph states corresponding to directed weighted graphs have been studied. These states have been constructed with the action of $RXX(\phi_{ij})$ gates on an arbitrary separable multi-qubit state (\ref{eq:2:1}). We have found the geometric measure of entanglement of a qubit with other qubits in the quantum graph state corresponding to arbitrary weighted directed graphs.

We have obtained the relation of the entanglement of the quantum graph states with the graph properties. Specifically, in the general case, the entanglement of a qubit $q[k]$ with other qubits in the quantum graph state is related to the weights of ingoing and outgoing arcs with respect to the vertex $k$ representing it in the graph, as well as the sum of weights of antiparallel arcs corresponding to the vertex (\ref{E_gen}). In the particular case of the absence of bidirected arcs in the graph, and when the weights of arcs entering the corresponding vertex $k$ and the weights of arcs leaving $k$ are equal ($\phi_{mk}=\phi^{(in)}_k$, $\phi_{kn}=\phi^{(out)}_k$), the entanglement of qubit $q[k]$ in the graph state (\ref{eq:2:1}) is related to the outdegree and indegree of the corresponding vertex $k$ (see (\ref{entkk})). For a quantum graph state corresponding to an unweighted and undirected graph, the entanglement of qubit $q[k]$ in the graph state is related to the degree of the corresponding vertex (\ref{par2}).

Additionally, a quantum protocol for the study of the entanglement of quantum graph states (\ref{eq:2:1}) has been presented (see Fig. \ref{fig:1}). As an example, a quantum graph state corresponding to a chain has been considered. We have studied the dependencies of the geometric measure of entanglement of quantum graph states on its parameters analytically as well as through calculations on AerSimulator. The results of quantifying the entanglement are presented in Figs. \ref{fig:4}, \ref{fig:5}.

\section{Acknowledgments}

This work was supported by the Virtual Ukrainian Institute of Advanced Studies. The author thanks Prof. Tkachuk V. M. for many useful comments and support during the research studies

\end{document}